\documentclass{elsart}
\usepackage{graphicx}
\usepackage{amsmath}
\usepackage{amssymb}

\newcommand{\be}{\begin{eqnarray}}

\newcommand{\ee}{\end{eqnarray}}
\newcommand{\NN}{\nonumber \\}

\begin{document}

\begin{frontmatter}

\title{Impact Parameter Dependence in the Balitsky-Kovchegov Equation}

\author[RBRC]{Takashi Ikeda}
,
\author[RBRC,BNL]{Larry McLerran}

\address[RBRC]{RIKEN BNL Research Center, Brookhaven National Laboratory,
 Upton, NY 11973 USA}
\address[BNL]{Physics Department, Brookhaven National Laboratory,
 Upton, NY 11973 USA}

\begin{abstract} 
We study the impact parameter dependence of solutions to
the Balitsky-Kovchegov (BK) equation.  We argue that if the 
kernel of the BK integral equation is regulated 
to cutoff infrared singularities, then it can be approximated 
by an equation without diffusion in impact parameter.  For some purposes, when
momentum scales large compared to $\Lambda_{QCD}$ are probed, the kernel
may be approximated as massless. In particular, we find that the 
Froissart bound limit is saturated for physical initial conditions
and seem to be independent of the cutoff so long as the cutoff is 
sufficiently large compared to the momentum scale associated with the large
distance falloff of the impact parameter distribution.
\end{abstract}

\end{frontmatter}

\section{Introduction}

The issue of impact parameter dependence of parton distribution functions 
is an old one.  Recently, it has been possible to address this issue
quantitatively in the context of the BFKL and the Balitsky-Kovchegov
evolution equations \cite{bfkl}-\cite{k}.  One would like to compute the parton 
phase space density,
\begin{equation}
	{{dN_{gluon}} \over {dY d^2k_T d^2b}} 
= {1 \over \alpha_S} {{2 N_c} \over
{(2\pi)^2 \pi^2}} {{\phi(Y,k_T,b)} \over k_T^2}
\end{equation}
where $Y$ is the gluon rapidity, $k_T$ is its transverse momentum
and $b$ is the transverse spatial point where the gluon density is measured.
The quantity $\phi$ is related to its coordinate space analog by
\begin{equation}
	{{N_Y(\vec{r}_T,\vec{b})} \over r_T^2} = 
\int {{d^2k_T} \over {(2\pi)^2}} 
e^{i\vec{k}_t\cdot \vec{r}_t} {{\phi(y,\vec{k}_T,\vec{b})} \over k_T^2}
\end{equation}
In the Color Glass Condensate description of high density gluonic matter,
valid at small x \cite{mq}-\cite{jkmw},
\begin{equation}
	N_Y(\vec{r}_{\text{\tiny $T$}},\vec{b}) = {1 \over N_c} 
<Tr(1 -U^\dagger (\vec{x}) U(\vec{y}))>
\end{equation}
where $U$ is a line integral of the gluon field which goes over the
rapidities of the gluons, and where
\begin{equation}
	\vec{r}_{\text{\tiny $T$}} = \vec{x}_{\text{\tiny $T$}} - \vec{y}_{\text{\tiny $T$}}
\end{equation}
and
\begin{equation} 	
       \vec{b} = {{(\vec{x}_{\text{\tiny $T$}} + \vec{y}_{\text{\tiny $T$}})} \over 2}
\end{equation}
As $r_T \rightarrow 0$, $N \rightarrow 0$, and as $r_T \rightarrow \infty$,
$N \rightarrow 1$.  This follows because at large distances, the 
expectation values of $<U(x)U(y)> \rightarrow <U(x)><U(y)>$, and
$<U(x)> \rightarrow 0$ by gauge invariance.  This means that correlations
should fall off at long distances, and it means that isolated source of color
charge cannot propagate in QCD.

The distribution function $N$ has been argued to satisfy a non-linear evolution
equation, the Balitsky-Kovchegov (BK) equation.  It is usually
written in terms of the variables $x,y$ rather than a relative coordinate
and impact parameter $r,b$. (Below, we drop the arrow and subscript $T$
representing the transverse variables.)
\be
{\partial \over {\partial Y}} N_Y(x,y)  &=&  
\overline \alpha \int d^2z {{r^2 } \over {(x - z)^2
(z - y)^2}} \bigl( N_Y(x,z)  + N_Y(z,y)  \NN
& & \hspace{4cm} - N_Y(x,y) - N_Y(x,z) N_Y(z,y) \bigr)
\label{bk}
\ee
where 
\begin{equation}
	\overline \alpha = {{\alpha_S N_c} \over {2 \pi^2}}
\end{equation}

Let us see what happens if we try to find a solution to this equation which
has exponentially falling behavior at large impact parameter.  Let us take
as initial condition for this equation\cite{numbfkl}
\begin{equation}
	N_0(r,b) = 1-exp\left( -f(r)e^{-2\mu b} \right)
\label{ansatz}
\end{equation}
In this equation, $\mu$ should be taken to be $m_\pi$ for the long distance 
falloff, since this should be controlled by isosinglet exchange, and the 
2 pion state is the lowest energy strongly interacting state with these
quantum numbers.

At large impact parameters, for fixed $r$, this goes into the factorized
form
\begin{equation}
	\lim_{b \rightarrow \infty} N_0(r,b) = f(r) e^{-2 \mu b}
\label{dist_fact}
\end{equation}
On the other hand, if $f(r)$ increases as $r$ increases, then for some
large enough $r$, the exponential on the right hand side of 
Eq. (\ref{ansatz}) becomes small, and $N \rightarrow 1$,
its saturated value.  This is the behaviour one expects from the operator
definition of the distribution function $N$, since we expect that
\begin{equation}
	<U(r+R)U^\dagger (R)>
\end{equation}
should approach zero at large $r$.  This requires that $f(r)$ be singular
at large $r$. It is also true that if we make $f$ grow too rapidly as
$r \rightarrow \infty$, we may introduce a stronger dependence on the 
infrared than desired.  In what follows, we will choose initial conditions
so that $f$ goes not faster than $r^2$ as $r \rightarrow \infty$. 
This constraint on the growth at large $r$ is not spoiled by evolution, 
as we will see numerically below.
Specifically, we shall choose
\begin{equation}
f(r) = \frac{(c r)^2}{1+c r} \, .
\label{eq:hybrid_init}
\end{equation}
Here, $c$ is the mass-dimensional constant corresponding to the initial
saturation scale at zero impact parameter.

Let us see if the generic features of this form can survive evolution in
rapidity, according to the BK equation.
Let us consider $N_Y(x,y)$ for $x = R$ and $y  = 0$ where $R \gg 1/\mu$.
After one iteration in time (rapidity), the contribution $N_Y(z,y)$ in
the BK integral is big for $z \sim 0$.
The kernel is of order  1.  This generates a non-exponentially falling
contribution.  Next if we iterate again, and look at $N_Y(x,y)$ for
$ x \sim y \sim R$ ($r \sim 1/\mu$ and $b \sim 2R$), we see that 
$N \sim 1/R^4$.  This means that the solution evolves towards a power
law falloff in the impact parameter, not an exponential as one would have thought
from the initial conditions.

The basic reason that the BK equation generates a bad behaviour in impact 
parameter is because of the massless nature of the kernel of the BK equation.
The point of this paper will be to show that if one regulates the kernel
by introducing an infrared cutoff, then the exponential falling nature of 
BFKL which may be built into an initial condition will be maintained by 
evolution.

What we will argue is that the BK equation may be replaced by an equation
which has no diffusion in impact parameter.  This equation is well defined
with no infrared cutoff in the kernel.  The corrections to this equation which
involve diffusion remain small so long as the infrared cutoff is
larger than or equal to $\mu/2$, the scale
associated with the impact parameter falloff
and so long as one measures the distribution functions at momentum
scales $Q^2 \gg \Lambda^2_{QCD}$ at not too large impact parameters.  
Roughly speaking, $Q \sim 1/r$, so one
probes the short distance structure of $f(r,b)$.

This form of the evolution equation is  sufficient to establish the 
Froissart bound for high $Q^2$ processes \cite{Froissart}-\cite{fiim}.  
It is reasonable to assume 
that this bound is also true for total hadronic cross sections where
$Q^2 \sim \Lambda_{QCD}^2$.  
We find that the Froissart bound is saturated, and the 
scale associated with the cross section reflects the initial conditions,
not the scale in the cutoff, so long as the scale in the cutoff is greater than or
equal to one half that scale of 
falloff in the initial impact parameter profile. 
\section{Regularizing the BK Kernel}

Let us try the simplest possible modification of the BK kernel in 
Eq. (\ref{bk}). We define
\begin{equation}
	K(\lambda, x,y,z)= {{(x-y)^2} \over {(x-z)^2 (z-y)^2}}
e^{-\lambda \mid x-z \mid} e^{-\lambda \mid z-y \mid } \, .
\end{equation}
Here, the parameter $\lambda$ represents the non-perturbative mass scale.
We will take $\lambda$ as a parameter in our analysis below.
The smallest physically acceptable value for $\lambda$ we believe is
$\lambda = m_\pi$.  It might be larger, since in the evolution
of the distribution function, the growth of distribution functions
might be associated with vector meson exchange, as is true in some models.
The generic conclusions we will draw below rely more on $\lambda$
being larger than $\sim m_\pi$, than on its specific value. 
In the limit that $\lambda \rightarrow 0$, this is the kernel of the BK
equation in Eq. (\ref{bk}).

To understand how the infrared cutoff in the kernel above
preserves the features of our ansatz for the initial condition for 
$N$, we write the BK equation with the modified kernel in terms of  
relative coordinates
\be 
&&{\partial \over {\partial Y}} N_Y(r,b)  =  {\overline \alpha} \int d^2z
{r^2 \over{ \mid {r \over 2}  -z \mid^2 \mid {r \over 2}  +z \mid^2}}
e^{-\lambda \mid {r \over 2}  -z \mid }
e^{-\lambda \mid {r \over 2}  +z \mid } \NN 
&&\times \left( N_Y\bigl({r \over 2}  -z, {1 \over 2} (z+{r \over 2}) +b\bigr)
 + N_Y\bigl({r \over 2}  +z, {1 \over 2} (z-{r \over 2}) +b\bigr)\right.  \NN
&&  \left.  
 -N_Y(r,b)   - N_Y\bigl({r \over 2}  -z, {1 \over 2} (z+{r \over 2}) +b\bigr) 
  N_Y\bigl({r \over 2}  +z, {1 \over 2} (z-{r \over 2}) +b\bigr) \right)  .
\label{bkrb}
\ee

Let us check how such an ansatz for the kernel affects the solution.  We 
assume the initial condition is given by the ansatz of Eq. (\ref{ansatz}).

\begin{figure}
\begin{center}
\includegraphics[width=0.50\textwidth]{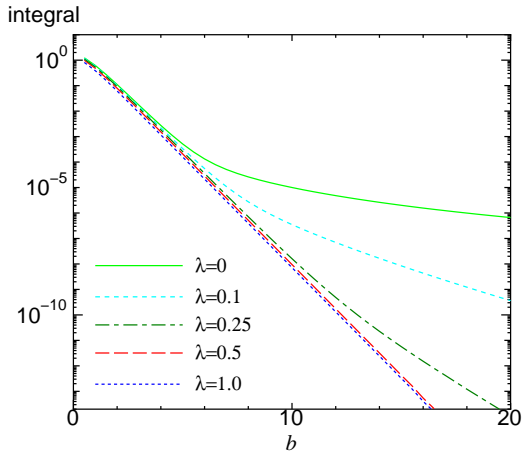}
\caption{\label{fig:reg_kernel} Impact parameter dependences of
 (\ref{eq:fullbk_integ}) for various $\lambda$ are
 plotted for $r = 0.1$ at $\cos \theta =0$. Solid, short dashed,
 dot-dashed, dashed and dotted lines correspond to those for
 $\lambda=0,0.1,0.25,0.5$ and $1$.}
\end{center}
\end{figure}
     
First consider what happens if we keep $r$ finite and go to large $b$.
Unless $\vec z \sim \pm  2\vec{b}$, the functions $N$ in the BK
equation are exponentially small.  If this is satisfied, then 
the kernel of the equation is of order $e^{-4 \lambda b}$ 
so that we preserve the form 
of the solution so long as $\lambda \ge \mu/2$ because the
distribution in the kernel is factorized in Eq. (\ref{dist_fact}).
(We expect that $\mu \ge m_\pi$ for the large distance asymptotics
of the regulated BFKL kernel, as noted in Sec. 1.  Therefore this
constraint is a rather weak one.)
In order to be precise, we evaluate the following integral
corresponding to the right-hand side of the BK equation
\be
&&\int d^2z \frac{(x-y)^2}{(x-z)^2 (z-y)^2} e^{-\lambda |x-z|}
e^{-\lambda |z-y|} \NN
&& \hspace{2cm} \times \bigl( N_0(x,z) + N_0(z,y) - N_0(x,y) - N_0(x,z)N_0(z,y)\bigr) \, ,
\label{eq:fullbk_integ}
\ee
with the initial condition (\ref{ansatz}) together with
(\ref{eq:hybrid_init}). Throughout this paper, we use the unit $\mu \equiv 1$ and set
$c=8.0$ in (\ref{eq:hybrid_init}) in numerical calculations.
Figure \ref{fig:reg_kernel} shows the impact parameter dependence of this integral
for dipole size $r=0.1$ and orientation 
$\cos \theta \equiv \vec{r} \cdot \vec{b}/rb = 0$. Here, we have assumed
cylindrical symmetry and dropped the dependence on the azimuthal angle.
This clearly shows that, while the power law tail is generated for
$\lambda = 0$, the exponential falling tail in the initial condition can
be maintained for $\lambda \ge \mu/2$. This power law tail can be seen in
Fig. 7 of \cite{Golec-Biernat:2003ym} where the same integral with no
infrared cutoff has been evaluated for the initial condition of the Glauber-Muller
form with a steeply falling profile in $b$, that is, $e^{-b^2}$.
Therefore, the massless  property of the BK kernel, not the form of the
(initial) distribution, may originate the power law tail in impact
parameter.  For $\lambda \le \mu/2$, the behaviour of the 
integral can be shown to be $e^{-4 \lambda b}/b^4$, from the structure
of the integral, and this fits the observed distribution well.

Now consider what happens if we take the limit that $r \rightarrow \infty$
with $b \gg 1/\mu$, but $b$ fixed.  In this case, the kernel vanishes,
and the distribution function maintains its original form, as seen in
numerical solutions below.  This is good
because in this region there is little matter, but one is in the 
non-perturbative region.  On general grounds, the distribution function
should tend to one in this region, and should not be much affected by evolution
due to the presence of the Color Glass Condensate.

Note also that if we look at the problem term which generated long distance 
singular behaviour in our original analysis of the massless unregulated kernel,
we need to look at $r \sim R$ and $b \sim R/2$.  The potentially
dangerous region comes from  $z \sim -R/2$.  Then this
contribution is suppressed by $e^{-\lambda R}$ from the regulated kernel
which is small so long as $ \lambda \ge \mu/2$ as was the case above.

To summarize, the modification of the kernel we propose does in fact
have the correct properties to preserve the behavior we expect of
$N_Y(r,b)$ based on general principles.

\section{An Equation Ignoring Impact Parameter Diffusion}

\subsection{Discussion}
The Balitsky-Kovchegov equation as written in the $r,b$ basis as in 
Eq. (\ref{bkrb}) generates a power law tail in inverse impact
parameter if the kernel
is not regulated.  On the other hand, the regulated kernel generates no such
tail.  This suggests that if we compute the distribution function $N(r,b)$
on scales of $r$ which are small compared to the scale of variation in $b$,
then one may be able to ignore impact parameter diffusion.  

Looking at Eq. (\ref{bkrb}), we see that if the exponential tail
of the impact parameter cutoff satisfies $\lambda \ge \mu/2$, then the 
dominant contribution in the integration over z comes from $z \sim r$.
In the region where $z \sim 1/\mu$, we get a contribution of order $r^2 \mu^2$.
Clearly, ignoring impact parameter diffusion should be valid so long as we have
a cutoff kernel.

In practical terms, this means that a good lowest order approximation should
be generated if we expand the terms inside Eq. (\ref{bkrb})
\begin{equation}
	N(r^\prime, b^\prime ) = N(r^\prime, b + \Delta b)
\end{equation}
in powers of $\Delta b$, and keep the first non-leading terms.  The lowest 
order term in this gives an equation local in impact parameter
(after transforming $z \to z - \frac{1}{2}r$)\cite{levin}
\be
	{\partial \over {\partial Y}} N_Y(r,b)& = & 
\overline \alpha \int {d^2z} 
{r^2 \over{ \mid r -z \mid^2 \mid z \mid^2}}
e^{-\lambda \mid r -z \mid }
e^{-\lambda \mid z \mid } \bigl( N_Y(r -z,b)  \NN
& & \hspace{1cm}  + N_Y(z,b) - N_Y(r,b) - N_Y(r -z,b)N_Y(z,b) \bigr) \, .
\label{bknodb2}
\ee
This equation clearly preserves the initial conditions at large impact
parameter.  If $r \ll 1/\mu$, it also has the property that one can ignore
the $\lambda $ dependence in the evolution equation, so long as
$\lambda \ge \mu/2$, as we shall assume.
This means the solution to this equation has all the properties assumed
in Ref. \cite{fiim}, where the Froissart bound was argued to be true when
measurements were made at scales $Q^2 \gg \mu^2$.  

The solution to this equation should be roughly of the following form:
In the limit where the distribution function is itself small
and where $r \ll 1/\lambda$, the solution is
the ordinary $BFKL$ solution of the linear $BFKL$ evolution equation
times an impact parameter profile.  This solution will hold until
the function $N$ becomes of order 1.  This should occur roughly when
\begin{equation}
	(rQ_{sat}(Y))^\gamma e^{-\mu b} \sim 1 \, ,
\end{equation}
or
\begin{equation}
	r \sim \frac{e^{\mu b/\gamma}}{Q_{sat}(Y)}
\end{equation}
where $Q_{sat}(Y)$ is the saturation momentum  of the ordinary BK equation.
For rapidities beyond the critical rapidity for which the amplitude
$N_Y(r,b)$ first becomes of order 1, the amplitude remains of order 1 
for larger rapidity because the BK equation doesn't violate the unitarity.

On the other hand, when  $r \ge 1/\lambda$, the kernel
of the BK equation introduced above cuts off further evolution, 
and the function $N_Y(r,b)$ is frozen to its initial condition.  
  
It is difficult to get a direct computation of the correction to the
zero impact parameter diffusion approximation.  Naive expansion in a Taylors
series around zero impact parameter diffusion leads to an equation with mild
infrared singularities.  Nevertheless, the estimate made above, that these 
corrections should be of order $r^2 \mu^2$, up to logarithms, is still true.
This of course requires that $\lambda \ge \mu/2$.  If this was not true, the
tail of the distribution would be modified in an unmanageable way.  
Nevertheless, the leading order solution is valid so long as 
we study small $r$, and does not requires that $\lambda \ge \mu/2$.  
On the other hand, at large $r$, our form of 
the kernel has the physically plausible behavior that $N_Y(r,b)$ does
not evolve in $Y$.

\subsection{Numerical Solutions}

In this section, numerical solutions of Eq. (\ref{bknodb2}) for 
$\lambda = 0$ and $1$ are calculated. We used the same
parameters $\alpha_S=0.2$, $N_c=3$ and $c=8$ in the initial condition
(\ref{ansatz}) together with (\ref{eq:hybrid_init}), as in Sec. 2. 
Since Eq. (\ref{bknodb2})
is local in $b$ and our initial
condition depends only on $r (\equiv |\vec{r}|)$ and $b (\equiv |\vec{b}|)$
but not on the angle $\theta$ between $\vec{r}$ and $\vec{b}$, 
the solution $N_Y$ at any
rapidity has the same property, that is, $N_Y(\vec{r},\vec{b}) = N_Y(r,b)$.

The numerical method to solve Eq. (\ref{bknodb2}) is similar to the one
for the solutions of the full BK equation \cite{Golec-Biernat:2003ym},
and highest rapidity in our calculations is 40.

\subsubsection{The dipole size dependence of the solution}

\begin{figure}
\begin{center}
\includegraphics[width=0.80\textwidth]{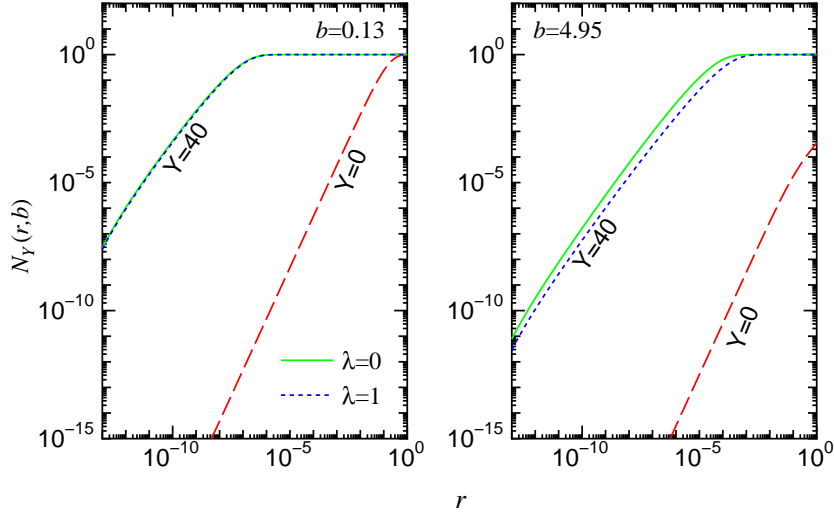}
\caption{\label{fig:dist_rdep} The $r$ dependence of distribution
 functions for $b=0.13$ (left panel) and $4.95$ (right
 panel). Dashed lines represent the initial distribution. Solid and
 dotted lines correspond to the distributions for $\lambda=0$ and $1$ at
 $Y=40$. }
\end{center}
\end{figure}

First, let us examine the rapidity evolution of the distribution
function for small dipole sizes $r < 1$. In Fig. \ref{fig:dist_rdep}, the
distribution functions for $\lambda = 0$ and $1$ are plotted as 
a function of $r$. Left and right panels correspond to the distributions
at $b=0.13$ and $4.95$. In each panel, the dashed line is the initial
distribution, and solid and dotted lines are the distributions for
$\lambda=0$ and $1$ at highest rapidity $Y=40$. While the distribution 
functions for $\lambda = 0$ and $1$ are in good agreement for small 
$b (=0.13)$ as shown in the left panel, the evolution for $\lambda=1$ 
is slightly suppressed for large $b (=4.95)$ in the right panel.
We also found the exponent for small $r$ changes from
$r^2$ in the initial distribution to $r^{2 \gamma_0}$ with 
$\gamma_0 \sim 0.6-0.7$ at high rapidities, 
independently of $b$ and $\lambda$ \cite{iim}-\cite{mt}. 
This $\gamma_0$ is called anomalous
dimension and the value is consistent with other studies \cite{fiim}.

\begin{figure}
\begin{center}
\includegraphics[width=0.50\textwidth]{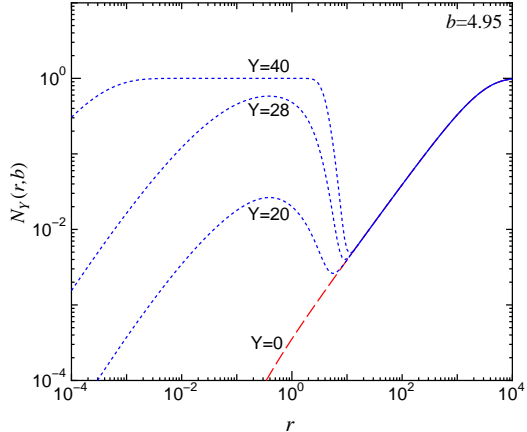}
\caption{\label{fig:dist_rdep_large} The distribution function at 
$b=4.95$ for $\lambda=1$ 
is plotted as a function of $r$ at various rapidities.}
\end{center}
\end{figure}

Next, let us examine how the regulated kernel affects the evolution of
large dipole size. Figure \ref{fig:dist_rdep_large} shows the
distribution function at $b=4.95$ for $\lambda=1$ including the region 
$r> 1/\lambda$ where the regulated kernel may affect the evolution.
It is observed in this figure that the evolution of the distribution
function is suppressed around $r \sim 1 (=1/\lambda)$ and the
distribution is frozen to its initial value in the region $r \gg 1$, 
as we mentioned in the previous section.
If the initial distribution reaches to its
saturated value in the region $r \gtrsim 1/\lambda$, it doesn't evolve
any further in such  region and the regulated kernel doesn't affect its evolution. 
This is the case for small $b$ in our initial condition, as you can see
in the left panel of Fig. \ref{fig:dist_rdep}. 

That said, the dip shown in Fig. 3 for large $r$ is in an unphysical region
where the separation of the sources of color charge within the dipole
are at unphysical size scales.  This is a region dominated by non-perturbative
physics, and ascribing physics to this region is no doubt problematic.
This region causes no problem in the solution of the equation we consider,
and the dip is presumably an artifact of an incomplete non-perturbative
treatment of the correlation function.

\subsubsection{The impact parameter dependence of the solution}
        
\begin{figure}
\begin{center}
\includegraphics[width=0.80\textwidth]{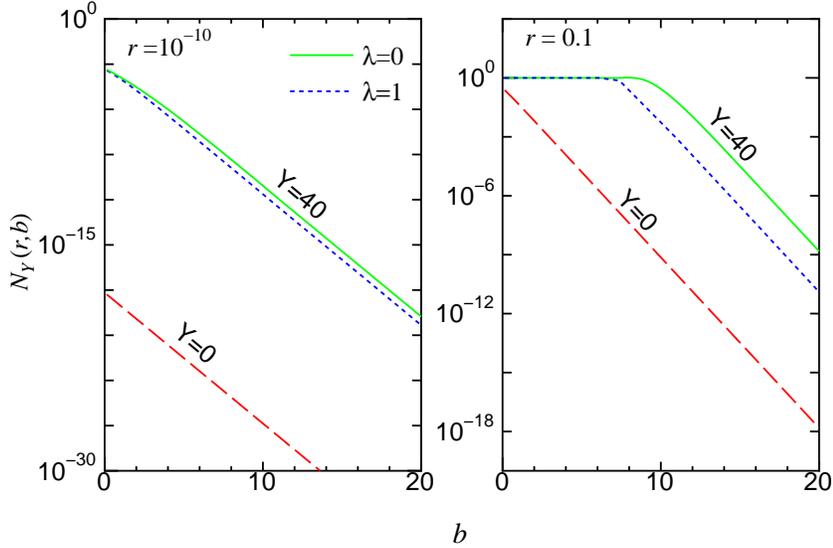}
\caption{\label{fig:dist_bdep} The distribution
 functions for fixed $r$ at $Y =0$ and $40$ are plotted as a function
 of $b$. Left and right panels correspond to the distributions for 
 $r = 10^{-10}$ and $0.1$. Dashed lines represent the initial distribution. Solid and
 dotted lines correspond to the distributions for $\lambda=0$ and $1$.}
\end{center}
\end{figure}

In Fig. \ref{fig:dist_bdep}, the impact parameter dependences of the distributions at fixed 
$r$ are plotted for $Y=0$ and $40$. It is clearly observed that the
exponential falling behavior in the initial condition is maintained at
high rapidities both for $\lambda=0$ and $1$, as we expected above.
While, for small dipole size $r=10^{-10}$, the distribution functions
for $\lambda=0$ and $1$ at high rapidities are in good agreement,
the evolution for $\lambda=1$ is a little suppressed compared to that
for $\lambda=0$ for $r=0.1$ due to the regulated kernel.

\subsubsection{Saturation scale}

Next, let us examine the rapidity evolution of the saturation
scale $Q_{sat}(Y)$. We define $Q_{sat}(Y,b)$ as an inverse of 
the lowest $r$ satisfying
\begin{equation}
N_Y(r = 1/Q_{sat}(Y,b),b) = \kappa \, ,
\label{eq:def_qs}
\end{equation} 
where $\kappa$ is order of unity. 
The explicit value of $\kappa$ doesn't matter and $\kappa$ is set to 1/2
below. This definition is the same as in Ref. \cite{Golec-Biernat:2003ym}.

\begin{figure}
\begin{center}
\includegraphics[width=0.50\textwidth]{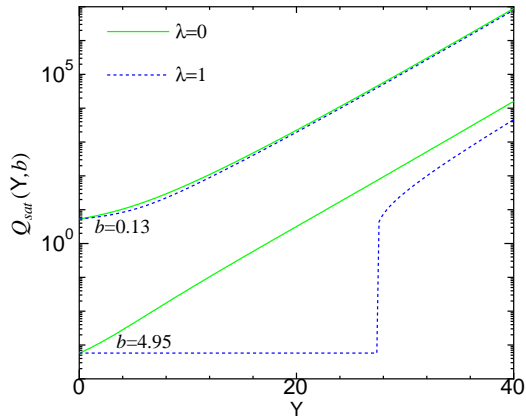}
\caption{\label{fig:qs_ydep} The saturation
 scale for $\lambda = 0$ (solid line) and $1$ (dotted line) are plotted
 as a function of rapidity. Upper and lower lines correspond to $Q_{sat}(Y,b)$ for
 $b  = 0.13$ and $4.95$.
}
\end{center}
\end{figure}
   
Figure \ref{fig:qs_ydep} shows the saturation scale for 
$b  = 0.13$ and $4.95$ as a function of rapidity. As you can see,
the rapidity evolutions with $\lambda = 0$ and $1$ almost coincide for
small $b$ (=0.13 in this figure), in other words, the infrared cutoff doesn't change the 
evolution of the saturation scale for the small impact parameter. 

For large $b (=4.95)$, the saturation scale for $\lambda=0$ evolves
similarly to that for $b=0.13$. 
On the other hand, the $Q_{sat}$ for $\lambda = 1$ stays in its initial 
value until the rapidity goes beyond a certain value, and then starts
evolving. Looking carefully at Fig. \ref{fig:dist_rdep_large}, we can understand
this behavior. As rapidity increases, the height of the bump of the distribution for
$\lambda=1$ around $r \sim 1$ also increases. Until the height of the bump reaches to
$\kappa$ in (\ref{eq:def_qs}), the $Q_{sat}(Y,b)$ is frozen to its
initial value.  (We should note that nothing particularly
exciting is happening with the correlation function itself when
$Q_{sat}$ makes a jump, and the jump is no doubt an artifact of
its definition, rather than corresponding to some rapid change in a physical
quantity.) 
When the height reaches to $\kappa$ (at $Y \sim 28$ for
$b=4.95$), the $Q_{sat}$
jumps to the lowest $r$ satisfying (\ref{eq:def_qs}) from the initial
value $1/Q_{sat}(Y=0)$ and then evolves normally because this is the
region  $Q_{sat}(Y,b) > \lambda$, 
{\it e.g.}, $r< 1/\lambda$ and the evolution of the distribution is not
affected by the regulated kernel. The point at which the $Q_{sat}(Y,b)$
jumps from the initial value and starts evolving is a monotonically
increasing function of $b$. For example, the $Q_{sat}$ at $b=3.3 (4.95, 7.4)$ has a
jump at $Y\sim 20 (28, 40)$, and for $b>7.4$, the $Q_{sat}$ for $\lambda=1$ is still frozen to
its initial value because of the highest rapidity $Y=40$ in our calculations.

It is noted that the slopes at high rapidities
are almost same independently of $b$ and $\lambda$, which means 
the saturation scale at high rapidities can be factorized in the form
\begin{equation}
Q_{sat}(Y,b) \sim exp(2 \pi \lambda_s \bar{\alpha} Y) \times N(b) \, ,
\end{equation} 
where $N(b)$ is an impact parameter profile function.
In our calculation, $\lambda_s$ is estimated around $2.1$ which is
consistent with that from the full BK equation
\cite{Golec-Biernat:2003ym}. 
This might mean that the rapidity evolution of $Q_{sat}(Y,b)$ is not
sensitive to the impact parameter diffusion.

\begin{figure}
\begin{center}
\includegraphics[width=0.50\textwidth]{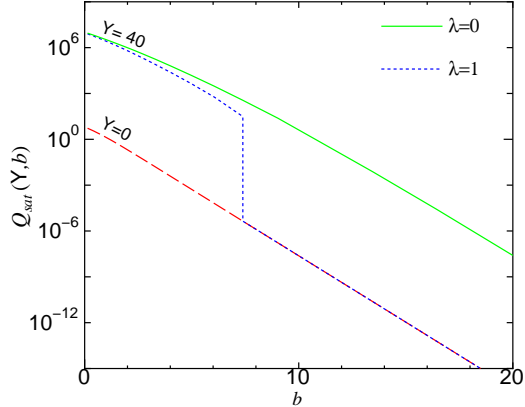}
\caption{\label{fig:qs_bdep} The impact parameter dependence of the saturation
 scale for $\lambda = 0$ (solid line) and $1$ (dotted line) is plotted
 at $Y=0$ and $40$. 
}
\end{center}
\end{figure}

In Fig. \ref{fig:qs_bdep}, the $b$ dependence of the saturation scale for
$\lambda=0$ and $1$ is plotted at $Y=0$ and $40$. For
small $b$, especially $b < 1$, the impact parameter dependences for
$\lambda =0$ and $1$ agree well. 
This is because the evolution of the distribution function for small $b$
doesn't change by the vacuum property of the regulated kernel, as seen in
the left panel of Fig. \ref{fig:dist_rdep}. 

We also found the exponential tail of the saturation scale for
$\lambda=0$ is maintained at high rapidities, {\it e.g.}, $e^{-2 b}$.
Of course, this exponential tail is totally different from that from the 
full BK equation without infrared cutoff of the kernel,
which reads the power-like tail $Q_{sat}(Y,b) \sim 1/b^{\gamma}$ with
$\gamma = 1.6-2.0$ at high rapidities \cite{Golec-Biernat:2003ym}. 
The exponential tail is preserved only when the kernel of
the BFKL equation is properly regulated in the infrared.

\section{The Growth of Black disc radius and The Froissart bound}

One of the most important issues in the saturation physics is how fast
the black disc radius in which $N_Y$ saturates grows in rapidity. This
is closely related to the problem of the Froissart bound
\cite{Froissart} which tells us that the total hadronic cross section is
bounded by
\begin{equation}
\sigma = \frac{2 \pi}{m_{\pi}^2} \ln^2(s/s_0) \, ,
\label{eq:fb}
\end{equation}
with energy $s \sim e^Y$.
We define the black disc radius $R_{BD}(r,Y)$ as
\begin{equation}
N_Y(r,b=R_{BD}(r,Y)) = \kappa \, ,
\end{equation}
where $\kappa$ is order of unity and $\kappa = 1/2$ is chosen as in the
case of the saturation scale.  

In the saturation regime, the total dipole-nucleus cross section 
is given by integrating the scattering
probability $N_Y$ over the impact parameter,
\begin{equation}
\sigma(r,Y) = 2 \int d^2b N_Y(r,b) \sim 2 \pi R_{BD}^2(r,Y) \, .
\end{equation}
If this $R_{BD}$ grows at most linearly with the rapidity $Y$, 
the Froissart bound is saturated.
Let us discuss this below.

\begin{figure}
\begin{center}
\includegraphics[width=0.49\textwidth]{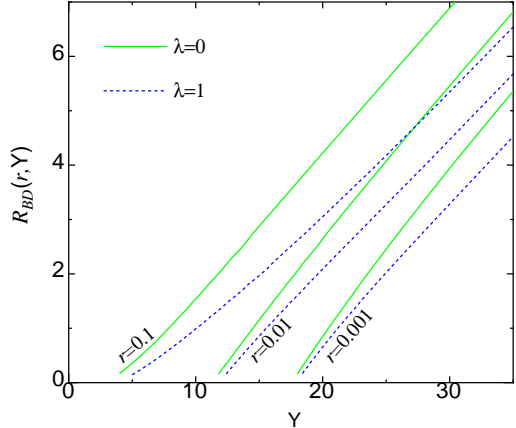}
\caption{\label{fig:Rbd_ydep} The black disc radius
 for $\lambda = 0$ (solid lines) and $1$ (dotted lines) is plotted
 as a function of rapidity at $r = 0.1, 0.01$ and $0.001$ from the 
 left to right.
}
\end{center}
\end{figure}

Figure \ref{fig:Rbd_ydep} shows the rapidity dependence of
the black disc radius for various dipole sizes. As we can see, in both cases for
$\lambda = 1$ and $0$, $R_{BD}$ grows linearly as a function of
rapidity. The slope at high rapidities for $\lambda=0$ is slightly larger than that for
$\lambda=1$ which doesn't depend on the dipole size if it is
small enough ($r<1$).
One might think it's a natural consequence of the local approximation ignoring
the impact parameter diffusion.
The important point, however, is that this linear growth of the black disc radius
in $Y$ is expected in the full BK evolution equation if the kernel of the equation
is regulated by the infrared cutoff, although the naive BK equation shows
the exponential growth and violates the Froissart bound.

\section{Summary}

The origin of the infrared cutoff in our analysis needs some understanding 
beyond what we present here.  It is clear that the lower bound we have
is a weak one and that the generic feature of Froissart bound
saturation appears to be insensitive to details of this
cutoff.  It is however not clear to us what sets the scale
of this cutoff.  Is it $2 m_\pi$, or $2 m_\rho$?  Might one derive an
effective meson theory which can generate a BFKL like
equation to describe this non-perturbative region?
Of course, it is purely non-perturbative and
cannot be explained by the perturbative QCD.

Although the considerations presented here are far from
rigorous, it seems that one does in fact have a simple intuitive 
picture of how the Froissart bound arises.  It is simply the
trade off between the exponential growth of the gluons as
a function of rapidity filling up and exponentially falling tail of
an impact parameter distribution set by initial conditions.

\section{Acknowledgements}
L.M gratefully acknowledges conversations with 
Edmond Iancu and J. Bartels on the subject of this talk.
T.I is much grateful to S.~Munier and A.~M.~Stasto for numerous valuable 
discussions.
T.I was supported through a Special Postdoctoral Researcher Program of RIKEN.
This manuscript has been authorized under Contract No. DE-AC02-98H10886 with
the U. S. Department of Energy.

\end{document}